\documentclass[11pt]{article}

\usepackage{amssymb, amsmath, stmaryrd, epsfig, fullpage,xspace} 
\def\mat#1{\ensuremath{#1}\xspace}

\newtheorem{thm}{Theorem}[section]
\newtheorem{theorem}[thm]{Theorem}

\newtheorem{corollary}[thm]{Corollary}

\newtheorem{lemma}[thm]{Lemma}

\newtheorem{prop}[thm]{Proposition}

\newtheorem{defin}[thm]{Definition}

\def\PRO#1{\begin{prop}#1\end{prop}}
\def\DEF#1{\begin{defin}#1\end{defin}}

\def\AR#1{\begin{align}#1\end{align}}

\def\SP#1{\begin{split}#1\end{split}}
\def\GA#1{\begin{gather}#1\end{gather}}

\def\CA#1{\begin{cases}#1\end{cases}}
\def\EQ#1{\begin{equation}#1\end{equation}}
\def\se#1{Sec.~\ref{#1}}

\def\de#1{Def.~\ref{#1}}
\def\defs#1#2{Defs.~\ref{#1} and \ref{#2}}
\def\pro#1{Prop.~\ref{#1}}

\def\re#1{Ref.~\cite{#1}}

\def\eq#1{Eq.\eqref{#1}}

\def\ens#1{\{#1\}}

\newcommand{\ket}[1]{\mbox{$|#1\rangle$}}
\newcommand{\gket}{\ket{\psi}}

\newcommand{\bra}[1]{\mbox{$\langle#1|$}}
\newcommand{\ketbra}[2]{\mbox{$|#1\rangle\langle#2|$}}

\newcommand{\p}{\ket{+}}

\newcommand{\tr}[1]{\ensuremath{\mathrm{Tr}(#1)}}

\def\den#1{\cD({#1})}

\def\M#1#2{{M}_{#2}^{#1}}
\def\tM#1#2#3#4{M_{#3#4}^{#1,#2}}

\def\cz#1#2{Z_{#1}^{#2}}
\def\cx#1#2{X_{#1}^{#2}}

\def\et#1#2{E_{#1#2}}

\def\ops#1{\llbracket#1\rrbracket_{op}}
\def\des#1{\llbracket#1\rrbracket_{de}}
\def\sem#1{\llbracket#1\rrbracket}
\def\nr{ \;|\; }
\def\given{\, \| \,}

\def\qc{\mathtt{qc}} 
\def\cc{\mathtt{c}} 
\def\qci{\mathtt{qc?}} 
\def\cci{\mathtt{c?}} 
\def\qco{\mathtt{qc!}} 
\def\cco{\mathtt{c!}} 

\def\rar{\rightarrow}

\def\lrar{\longrightarrow}

\def\Lrar{\Longrightarrow}

\def\Slar#1{\stackrel{#1}{\Lrar}}

\newcommand{\ox}{\otimes}

\newcommand{\ent}{\vdash}

\def\emptyset{\varnothing}
\def\bs{\backslash}

\def\mrm{\mathrm}
 
\def\mbf{\mathbf}

\def\scs{\scriptstyle}
\def\dps{\displaystyle}

\def\al{\alpha}

\def\ga{\gamma}
\def\Ga{\Gamma}

\def\la{\lambda} 
 
\def\sig{\mat{\sigma}}

\newcommand{\bfa}{\ensuremath{\mathbf{a}}\xspace}
\newcommand{\bfb}{\ensuremath{\mathbf{b}}\xspace}

\newcommand{\bi}{\ensuremath{\mathbf{i}}\xspace}

\newcommand{\bo}{\ensuremath{\mathbf{o}}\xspace}

\newcommand{\bA}{\ensuremath{\mathbf{A}}\xspace}
\newcommand{\bB}{\ensuremath{\mathbf{B}}\xspace}

\newcommand{\cA}{\mathcal{A}}

\newcommand{\cD}{\mathcal{D}}
\newcommand{\cE}{\ensuremath{\mathcal{E}}}

\newcommand{\cH}{\mathcal{H}}
\newcommand{\cI}{\mathcal{I}}

\newcommand{\cL}{\mathcal{L}}

\newcommand{\cN}{\ensuremath{\mathcal{N}}}

\newcommand{\cP}{\ensuremath{\mathcal{P}}}

\begin{document}

\title{Distributed measurement-based quantum computation}

\author{
Vincent Danos\\ Universit\'e Paris~7 \& CNRS\\ {\small\texttt{Vincent.Danos@pps.jussieu.fr}} 
\and Ellie D'Hondt\\Vrije Universiteit Brussel\\ {\small\texttt{Ellie.DHondt@vub.ac.be}}
\and Elham Kashefi\\ IQC - University of Waterloo\\ {\small\texttt{ekashefi@iqc.ca}}
\and Prakash Panangaden\\ McGill University\\ {\small\texttt{Prakash@cs.mcgill.ca}}
  }
\date{}  
\maketitle
  
\begin{abstract} 
We develop a formal model for distributed measurement-based quantum computations, adopting an agent-based view, such that computations are described locally where possible. Because the network quantum state is in general entangled, we need to model it as a global structure, reminiscent of global memory in classical agent systems. Local quantum computations are described as measurement patterns. Since measurement-based quantum computation is inherently distributed, this allows us to extend naturally several concepts of the measurement calculus~\cite{Danos04b}, a formal model for such computations. Our goal is to define an assembly language, i.e. we assume that computations are well-defined and we do not concern ourselves with verification techniques. The  operational semantics for systems of agents is given by a probabilistic transition system, and we define operational equivalence in a way that it corresponds to the notion of bisimilarity. With this in place, we prove that teleportation is bisimilar to a direct quantum channel, and this also within the context of larger networks. 
\end{abstract}

\section{Introduction}

Measurement-based models provide an intriguing new framework for thinking about quantum computation.  While quantum circuits are still widely considered as a convenient formalism for describing algorithms, using measurements to steer quantum computation is considered a serious alternative.  Due to their inherently probabilistic nature, measurements were long thought to be a disturbance to quantum computations -- unavoidable though they are when wanting to read out the final output of a computation.  That they can be an active component of a computation has been known for quite some time through the teleportation protocol.  Only much later was it realized that in fault-tolerant constructions, measurements can be quite useful.  Soon thereafter, with the advent of models such as the one-way quantum computer~\cite{Raussendorf02a} and the teleportation model~\cite{Nielsen03,Leung04}, it was established that measurements could not only be a recurring component of a computation, but the actual driving force behind it.  Moreover, the measurement paradigm throws a whole new light on the strategies for actual physical implementations of a quantum computer~\cite{Walther05}.

However, measurements are not the only crucial ingredient of these models: they are also inherently \emph{distributed}.  Indeed, it is the realization that a variation on the teleportation protocol not only transports but also \emph{transforms} quantum information, which is the basis of the teleportation model.  Likewise, the one-way quantum computer is all about transformation via measurement and transportation, this time by way of a generic entangled state, the \emph{graph state}.  One-qubit measurements on this state transform the logical qubits, i.e. the quantum inputs, while transporting them via a path of graph state qubits.  Again, non-local correlations provided by particular entanglement properties, together with measurements, steer the computation.  Of course quantum measurements remain intrinsically probabilistic, but this can be solved by applying corrections dependent on previous measurement outcomes, rendering computations effectively deterministic.  Note that measurement outcome dependencies are crucial in order to obtain universality of the model. As a result, the typically distributed notion of classical communication is also naturally present in these models. All of this is very nicely captured by the \emph{measurement calculus}~\cite{Danos04b}, a formal framework for one-way computations.  Measurement patterns are defined essentially by sequences of commands allowed in the one-way model.  From this one can define operational and denotational semantics and show their equivalence, and prove that notions of composition are well-defined.  More importantly, there is an associated \emph{rewrite system} which allows one to put any pattern into a standard form.  The measurement calculus, which can be seen as an \emph{assembly language}, proves to be a valuable tool for formal investigations into all measurement-based models; for example, one can easily show how the teleportation model reduces to the one-way model, via a conversion between the associated calculi~\cite{Danos05}.

Because of the inherently distributed aspect, measurement-based models for quantum computation are well-suited as a starting point for a formal model for distributed quantum computations.  By this we mean \emph{macroscopically} distributed, i.e. we are talking about coordinated actions between different parties.  Of these there are many examples in quantum computation~\cite{Nielsen00}: teleportation, of course, but also entanglement swapping, logic gate teleportation, cryptographic protocols, and also quantum versions of classical distributed applications such as leader election~\cite{DHondt04}.  However, a formal language for distributed quantum computation is lacking.  There have recently been interesting developments based on classical process calculi~\cite{Lalire04,Gay04}, which have focused mostly on the concurrency aspects. While the distributed nature of computations was introduced via types in \re{Gay04}, the aim there is to develop formal verification techniques. In this work, we take an assembly-language point of view, assuming that computations are well-defined. This results in a compact model, with which we can explore properties of distributed protocols, such as the coordination between agents. 

We define an assembly language for distributed applications, directly built on the most basic distributed model of all: the one-way quantum computer. We adopt a \emph{local} view and describe the system as a set of \emph{agents} communicating synchronously and operating on a globally entangled quantum state; this is explained in \se{defs}. In \se{asem} we develop a formal semantics for systems of agents as  probabilistic transition systems. Operational equivalence is defined in a way such that it corresponds to the notion of bisimilarity.  We then prove that quantum teleportation is bisimilar to a direct quantum channel, and this also within the context of other, possibly entangled, agents, in \se{televsqc}. While the correctness of teleportation has been proved within other formal frameworks before~\cite{Gay04,Abramsky04}, the bisimilarity approach, and specifically, taking into account larger contexts as well, is new. We conclude in \se{aconclusion}.

Some familiarity with the measurement calculus model is assumed; for an in-depth exposition we refer to \re{Danos04b}.

\section{Networks of agents}\label{defs}

The main concept in our model for distributed measurement-based quantum computations is that of an \emph{agent}.  Agents are localized processes which, executing concurrently, make up a distributed system.  Formally, we define agents in the following way.

\DEF{ 
An \emph{agent} $\bA(\bi,\bo): Q.\cE$, with  classical input $\bi$ and output $\bo$, and sort given by a set of qubit references $Q$, is defined by a finite event sequence $\cE$ composed of
 \begin{enumerate} 
 \item patterns command sequences $\cA$, with input qubits in $Q$;
 \item classical message reception $\cci x$ and sending $\cco y$, where $\cc$ is a classical channel, and $x$ and $y$ are names;
 \item  qubit reception $\qci x$ and sending $\qco q$, where $\qc$ is a quantum channel and $q$ a qubit reference. 
 \end{enumerate} 
An agent's \emph{state} is given by a classical environment $\Ga$, which is a partial mapping from names, i.e. classical variables and qubit references, to values.
\label{agents}
}

Notice that any pattern $\cP(V,I,O,\cA)$ trivially corresponds to agent $\bA:I.\cA$. In general the sort $Q$ equals $I\uplus I^{s}$, where $I$ is the local quantum input and $I^{s}$ are qubits of a shared entangled state supplied by the network -- see \de{network} below. The classical input $\bi$ and output $\bo$ allows us to model protocols such as superdense coding, in which an agent wishes to send classical values  to another agent with the help of a shared Bell state. The local state is used to store measurement outcomes resulting from local pattern executions, input from other agents over classical channels $\cc$, and classical input bindings from the local environment. Further bindings are added to the state as required; that is, for example, whenever a qubit is measured, only then the signal name $s_{i}$ corresponding to the qubit reference $q_{i}$ is added to the domain of $\Ga$ and bound to the classical measurement outcome $v$, denoted $\Ga[s_{i}\mapsto v]$. The classical output set $\bo$ determines which bindings in $\Ga$ have to be preserved for the final output of a computation. We denote the local state restricted to the classical output by $\Ga_{\bo}$. 

Our interpretation of agents is different than the usual process approach, in that agents in our setting always correspond to actual parties in a distributed network, denoted by the label \bA. Therefore, an expression of the form $\bA(\bi,\bo): Q.\cE$ should be read as: the agent with \emph{name} \bA (Alice) runs the \emph{program} $\cE$ with qubits and in- and outputs as specified. An agents thus is a piece of code running on a particular processor. In this context, it only makes sense to compose agents if the agent names are the same. We refer to this as agent composition, which is formally defined below. Note that, since we allow subsequent agent programs to supply extra classical and quantum inputs as well as further process outputs from a previous program, some care is required in determining the classical type and quantum sort of the resulting agent. Other than that, agent composition is just straightforward concatenation of event sequences. 

\DEF{
The composition of agents $\bA(\bi_{1},\bo_{1}): Q_{1}.\cE_{1}$  and $\bA(\bi_{2},\bo_{2}): Q_{2}.\cE_{2}$, which is only defined for agents with identical agent names, is denoted 
$$\bA[(\bi_{2},\bo_{2}): Q_{2}.\cE_{2}]\circ[(\bi_{1},\bo_{1}): Q_{1}.\cE_{1}]$$
and given by 
\EQ{
\bA(\bi,\bo):Q.\cE_{2}\cE_{1}\quad {with} \quad
\CA{
\bi=\bi_{1}\cup (\bi_{2} \bs \bo_{1})\\
\bo=\bo_{1}\cup \bo_{2}\\
Q=Q_{1}\cup (Q_{2}\bs Q'_{1})
} \text{  ,}
\label{typecomp}}
where $Q'_{1}$ is the output sort of the first agent.
\label{agentcomp}
}
Note that output sorts can be determined by inspecting an agent's event sequence. The general idea is that $\bo_{1}\subseteq \bi_{2}$ and   $Q'_{1}\subseteq Q_{2}$, so that subsequent programs are defined at least on the outputs of previous ones. This can always be arranged by assuming identity transformations for those names in $\bo_{1}$ and $Q'_{1}$ not appearing in $\cE_{2}$. In particular this means that $Q'=Q'_{2}$. Extra inputs in subsequent programs are encountered for example when several qubits are teleported one after each other. In this case, Alice needs to execute her side of the protocol several times in a row, supplying a new local quantum input every time she initiates the protocol. The same argument holds for classical input when sequencing several dense coding protocols. Of course this is best understood within the context of \emph{network} rather than agent composition, which is defined below. 

A network of agents consist of several agents executing their event sequence concurrently, together with a global shared entangled state. As the network quantum state is inherently nonlocal, there is no other option than to regard it as some kind of global memory -- even though we wish to adhere to a local view. This leads to the following definition.

\DEF{ 
A \emph{network of agents} $\cN$ is defined by a set of concurrently acting agents together with a shared quantum state, that is
\EQ{\SP{
\cN&= \bA_{1}(\bi_{1},\bo_{1}):Q_{1}.\cE_{1}\nr\dots\nr\bA_{m}(\bi_{m},\bo_{m}):Q_{m}.\cE_{m}\given \sig\\
&=|_{i}\bA_{i}(\bi_{i},\bo_{i}):Q_{i}.\cE_{i}\given \sig \text{  ,}
}}
where $\sig \in \den{\cH_{\uplus_{i}I_{i}^{s}}}$, with $Q_{i}=I_{i}\cup I_{i}^{s}$ for all $i$. 
\label{network}
}

The network state \sig in the definition is the initial entanglement resource which is distributed among agents. Local quantum inputs specified in $I_{i}$ are added to the network state \sig during initialization. In this way we can keep initial shared entanglement as a first-class primitive in our model. In this paper, we do not describe the actual procedure for producing \sig. Note that agents in a network need to have different names, since they correspond to different parties that make up the distributed system. In other words, concurrency comes \emph{only} from distribution; we do not consider parallel composition of processes in the context of one party. Finally, individual agents $\bA(\bi,\bo):Q.\cE$ trivially correspond to a network $\bA(\bi,\bo):Q.\cE\given\mbf{0}$. Therefore statements about networks affect individual agents as well. 

We define two different ways of composing networks of agents, namely sequential and parallel composition. Because of our interpretation of agents as distributed processes, there are some constraints on these operations. Sequential composition is only defined for networks containing the same agents; the idea is that agents carry out event sequences of both networks one after the other. Furthermore, agent composition must be defined as per \de{agentcomp}, that is, inputs and initial sorts of the second network must contain outputs and final sorts of the first. Formally, as follows.

\DEF{
The sequential composition of networks $\cN_{1}=|_{i=1}^{m}\;\bA_{i}(\bi_{1,i},\bo_{1,i}):Q_{1,i}.\cE_{1,i}\given \sig_{1}$ and $\cN_{2}=|_{i=1}^{m}\;\bA_{i}(\bi_{2,i},\bo_{2,i}):Q_{2,i}.\cE_{2,i}\given \sig_{2}$ is defined as
\EQ{
\cN_{2}\circ \cN_{1}=|_{i=1}^{m}\;\bA_{i}[(\bi_{2,i},\bo_{2,i}):Q_{2,i}.\cE_{2,i}]\circ[(\bi_{1,i},\bo_{1,i}):Q_{1,i}.\cE_{1,i}]\given\sig_{1}\ox \sig_{2} \text{  .}
}
\label{netcomp}
}

Note that we have overloaded the notation $\circ$ to denote both agent and network composition. As long as both networks have the same number of agents, one can always arrange for them two be sequentially composable by renaming agents. In fact one can even compose networks with a different number of agents by adding null agents, i.e. just agent names, to the network with fewer agents. The preparations of both networks are assumed to be defined on disjoint Hilbert spaces; the composed network's preparation is given by the tensor product of these. They are to be interpreted as different entanglement resources that are used by both networks. Parallel composition, which expresses that two networks are operating in parallel and independent of each other, is only defined for networks containing different agents. Again, one can always rename agents so that this is well-defined.

\DEF{
The composition of networks $\cN_{1}=|_{i=1}^{m}\;\bA_{i}(\bi_{1,i},\bo_{1,i}):Q_{1,i}.\cE_{1,i}$ $\given \sig_{1}$ and $\cN_{2}=|_{i=1}^{n}\bB_{i}(\bi_{2,i},\bo_{2,i}):Q_{2,i}.\cE_{2,i}\given \sig_{2}$ is defined as
\EQ{\SP{
\cN_{1}\ox \cN_{2}=&|_{i=1}^{m}\;\bA_{i}(\bi_{1,i},\bo_{1,i}):Q_{1,i}.\cE_{1,i} \\
&|_{i=1}^{n}\bB_{i}(\bi_{2,i},\bo_{2,i}):Q_{2,i}.\cE_{2,i}\\
& \|\;\sig_{1}\ox \sig_{2} \text{  .}
}}
\label{netcompp}
}

By combining sequential and parallel composition, one can express a broad range of network combinations. In \se{acomp} we show that the semantics of networks is preserved under these operations.

The definitions given in the above can be rendered concisely under the form of an abstract grammar. We use $[]$ instead of $|$ to separate choices for expressions, as the latter is already in use to denote parallel composition.

\EQ{\SP{
\cA &::= \mrm{nil} \;[]\; E \;[]\;M \;[]\;C \;[]\; \cA \ox \cA \;[]\; \cA.\cA\\
\cE &::=\cci x \;[]\; \cco x \;[]\; \qci q \;[]\;\qco q \;[]\; \cA \;[]\; \cE.\cE \\
\bfa &::= \bA(\bi,\bo): Q.\cE \;[]\; \bA[(\bi_2,\bo_2): Q_2.\cE_2]\circ[(\bi_1,\bo_1): Q_1.\cE_1]\\
\cN &::= |_{i}\bfa_{i}\given\sig \;[]\; \cN\circ\cN \;[]\; \cN \ox \cN
}}

In the above $E$, $M$ and $C$ stand for any entanglement, measurement or correction commands, and nil is the null command. We refrain from giving a grammar for names; rather, we use conventions for these described below. Notice that both pattern command sequences and networks can be viewed as processes in the traditional process algebra sense, with a composition operation transforming any two patterns, respectively networks, into a new pattern or network. The definition of event sequences, which can be composed sequentially, is also common. Agents, however, have a less clear status in the process algebra framework. This is exactly because they are constructs that formalize \emph{distributed} notions, and therefore there are constraints on how they may be composed. This is not a flaw of our model but rather a requirement if one wants to make distribution explicit.

There are several notational conventions we adhere to throughout this paper. When the context is clear, as is often the case, we do not explicitly mention inputs and outputs, but instead just write $\bA:Q$. Moreover, we do not write inputs, outputs, sorts, or preparations if there are none, sometimes writing $-$ for empty input or output sets. For example, $ \bA(-,\ens{x}).\cci x$ is an agent with no input and no qubits, while $ \bA(-,\ens{x}).\cci x \nr  \bB(\ens{y},-).\cco y$ is a network with no  preparation in which a classical value is exchanged between two agents. We write $\cci xy$ for $\cci x\cci y$ and similarly for other communicating channels. In \de{agents} we have used pattern command sequences rather than full patterns for brevity, with the convention that the input of a pattern is always given by those qubits in its computation space that belong to the agent's sort at the moment when the pattern is executed. The output of the pattern is simply given by those qubits that are not measured. For example, in the single-agent network $\bA:\ens{1,2}.\cx 4{s_1}\M01\et14$ qubit 1 is an input to the pattern and 4 is an output; we also write this as  $\bA:\ens{1,2}.\cH(1,4)$. Pattern qubits not in an agent's sort are assumed to be computation qubits initialized to $\p$ as before, and need not be mentioned explicitly in an agent's sort. On the other hand, agent qubits not mentioned in a pattern event, as qubit 2 above, are always assumed to be left alone. That is, we do not explicitly write the identity pattern $\cI$ applied to $\bA$'s remaining qubit $2$. We consider pattern command \emph{sequences} rather than singular commands so that we can use the big-step semantics of patterns. As can be seen from these examples, we often refer to qubits via numbers; specifically we mostly refer to qubit $q_{i}$ as qubit $i$, and $s_{i}$ instead of $s_{q_{i}}$ for the corresponding signal variable. This has the advantage that patterns look just as  they do in \re{Danos04b}. In what follows below, we use specific letters for specific names, in particular we use $q_{i}$ or just $i$ for qubit references, $x_{i}$ and $y_{i}$ for ordinary classical variables, $v_{i}$ for classical values, and $s_{i}$ for classical signal variables.

We subject networks of agents to definiteness conditions, which ensure that the computation is well-defined. 
\begin{description}
\item[(H0)] an agent's communication events operate only on qubits in its sort $Q$. Pattern events with computation space $V$ have inputs in $Q\cap V$;   
\item[(H1)] an agent's events depend only on values in its state $\Ga$;
\item[(H2)] each quantum and classical message reception event has a corresponding quantum, respectively classical message sending event;
\item[(H3)] all names, i.e. classical variables and quantum references, are unique.
\end{description}

\subsection{Operational semantics}\label{asem}

Before we provide concrete evaluation rules for distributed computations, we give some clarifying explanations and examples as to how execution proceeds in the local view.  Throughout network evolution, each agent has access to the network state $\sig$ via the qubits it owns, transforming $\sig$ whenever a pattern event is executed.  While these patterns are local, $\sig$ is \emph{not}, and to preserve all information on the correlations we need to keep \sig unreduced at all times. As an example, suppose that an agent $\bA$ owns the first two qubits of the system's state $\sig(1,2,3,4,5)$.  Its next event is to execute the Hadamard pattern $\cH (1,6)$ on its first qubit.  As stated above, we do not explicitly write the identity pattern $\cI$ applied to $\bA$'s remaining qubit $2$. More importantly, neither do we write explicit identity patterns for the qubits not belonging to $\bA$, in this case $3$, $4$ and $5$. In full, this means that we have an evaluation step as follows 
 \EQ{
  \sig, \bA:\ens{1,2}.\cH(1,6) \Lrar\sig', \bA:\ens{6,2} \text{  ,} 
  } 
where $\sig'=\sem{\cH(1,6)\ox\cI^{\ox 4}(2,3,4,5)}\sig$, that is, $\sig '=(H\ox I^{\ox 4}) \sig(H\ox I^{\ox 4})$.  We denote the transition relation  by $\Lrar$. Execution of $\cH$ occurs in a single transition step by relying on its big-step semantics. In this way we avoid getting into the actual details of pattern execution, which is not what this paper is about. Note especially that the sort of $\bA$ has changed.  This is because measurements are destructive, so that the input qubit $1$ has disappeared, and the output $6$ has taken its place.  Because of this, we sometimes explicitly write $\sig'(6,2,3,4,5)$ in the right-hand side of the above evaluation rule. The only remnant of $1$ is its corresponding measurement outcome $s_1$, which is recorded in the state $\Ga$, via the added binding $\Ga[s_{1}\mapsto v]$, where $v$ is the measurement outcome.  \bA  might subsequently send $s_{1}$ to other agents by an event $\cco s_{1}$,  so in general we cannot delete this entry from $\Ga$.  Essentially this means that when new qubit references are generated by the execution of subsequent local patterns, or by quantum communications from other agents, these names need to be \emph{unique}. Because we do not want to get into the actual details of this naming procedure, as we are considering our model to be at the level of an assembly language, i.e.  at a stage where naming conflicts have been resolved, we have imposed this as a definiteness condition in the above.

A final point concerns an agent's classical input and output, received from, respectively sent to, its own local system.  In the local view, pattern events can depend also on classical inputs, rather than only on measurement outcomes.  Because of the uniform structure of the local state $\Ga$ we can piggyback input dependencies onto the signal dependency structure of MC.  This is, as mentioned before, one of the reasons why MC is such a good basis on which to built a framework for distributed quantum computations. With these concrete examples in mind, we are now ready to develop the operational semantics of the local view. As usual, we first define rules in terms of  small-step transitions, after which we switch over to the big-step framework. 

The small-step transitions for distributed computations essentially describe how agents, and the network with them, evolve over different time steps.  We adopt a shorthand notation for agents, leaving out classical inputs and output, which do not change with small-step reductions.

\EQ{\SP{ 
\bfa_{i}&= \bA_{i}:Q_{i}.\cE_{i}\\
\bfa_{i}.E & = \bA_{i}:Q_{i}.[\cE_{i}.E] \\
\bfa^{-q}&=\bA:Q\bs\ens{q}.\cE\\ 
 \bfa^{+q}&= \bA:Q\uplus\ens{q}.\cE[q/x] \text{  ,}
}}
 
where $E$ is some event, and $\cE_{i}$ and $\cE'_{i}$ are event sequences.  A \emph{configuration} is given by the  system state \sig together with a set of agent programs, and their states, specifically
\EQ{ 
\sig, |_{i}\Ga_{i},\bfa_{i}=\sig, \Ga_{1},\bfa_{1}\nr \Ga_{2},\bfa_{2}\nr\dots \nr  \Ga_{m},\bfa_{m}\text{  .}
}
The small-step rules for configuration transitions, denoted $\Lrar$, are specified below; we give some explanations afterwards. When the system state is not changed in an evaluation step, we stress this by preceding a rule by $\sig \ent$. 

\GA{
 \frac{\sig, \cP(V,I,O, \cA) \lrar_{\la} \sig',\Ga'} {\sig,\Ga,\bA:I\uplus R.[\cE.\cP] \Lrar_{\la} \sig',\Ga\scs 
 \cup \dps \Ga',\bA:O\uplus R.\cE} \label{localop}\\
\notag\\
\frac{\Ga_{2}(y)=v}{\sig \ent ( \Ga_{1},\bfa_{1}.\cc?x \nr \Ga_{2}, \bfa_{2}.\cc!y \Lrar \Ga_{1}[x \mapsto v],\bfa_{1} \nr \Ga_{2},\bfa_{2})} \label{rendvcl}\\
\notag\\
 \frac{}{\sig \ent ( \Ga_{1},\bfa_{1}.\qc?x \nr \Ga_{2},\bfa_{2}.\qc!q \Lrar \Ga_{1},\bfa_{1}^{+q}\nr \Ga_{2},\bfa_{2}^{-q})} \label{rendvq} \\
\notag\\
 \frac{L \Lrar_{\la}R}{L \nr L' \Lrar_{\la} R \nr L'} \label{metarule}
}

Here, $\scs \cup$ denotes the union of outcome maps. Implicit in these rules is a sequential composition rule, which ensures that all events in an agent's event sequence are executed one after the other.  The first rule is for local operations; we have written the full pattern instead of only its command sequence here to make pattern input and output explicit. Because a pattern's big-step semantics is given by a probabilistic transition system described by $\lrar$, we pick up a probability $\la$ here.  Furthermore, an agent changes its sort depending on pattern's output $O$, as explained in the examples above. The next rule is for classical rendez-vous and is straightforward. For quantum rendez-vous, we need to substitute $q$ for $x$ in the event sequence of the receiving agent, and furthermore adapt qubit sorts. The last rule is a metarule, which is required to express that any of the other rules may fire in the context of a larger system. $L$ and $R$ stand for any of the possible left-, respectively right-hand sides of any of the previous rules, while $L'$ is an arbitrary configuration. Note that we might need to rearrange terms in the parallel composition of agents in order to be able to apply the context rule.  This can always be done since the order of agents in a configuration is arbitrary. In derivations of network execution, we often do not explicitly write reductions as specified by \eqref{metarule}, but rather specify which in which order the other rules fire in the context of the network at hand. It is precisely in this last rule that introduces nondeterminism at the network level, that is, several agent transitions may be possible within the context of a network at the same time. 

Starting from the small-step rules above we can now define the big-step semantics of a system of agents.  We first define computation paths, which run from initial to final configurations via small-step transitions. In the \emph{initial configuration}, all local states are given by the map containing the classical input bindings $\Ga_{\bi_{i}}$, while the network state is determined by the entanglement resource together with local quantum inputs. A \emph{final configuration} is one in which all agents have an empty command sequence, and in which all local states have been restricted to classical output bindings, $\Ga_{i,\bo_{i}}$. Because of the definiteness conditions we imposed each computation always ends with a final configuration.  Supposing the initial quantum state of agent $\bA_{i}$ is given by $\rho_{i}\in I_{i}$, paths are defined as follows. 
  
\DEF{ 
Given a network of agents $\cN=|_{i}\bA_{i}(\bi_{i},\bo_{i}):Q_{i}.\cE_{i}\given \sig $ and quantum inputs $\rho_{i}$, a \emph{path} $\ga$  is a \emph{maximal} sequence of configurations $\ens{C_{j}=\sig_{j},|_{i}\:\Ga_{i}^{j},\bfa_{i}^{j}\;, j=1,\dots,k-1}$, i.e. 
\EQ{\SP{ 
C_{1}&=\sig\ox_{i=1}^{m}\rho_{i}, |_{i}\:\Ga_{\bi_{i}},\bA_{i}:Q_{i}.\cE_{i}\\ 
C_{j}&\Lrar_{\la_{j}} C_{j+1}\\ 
C_{k}&=\sig_{k}, |_{i}\:\Ga_{i,\bo_{i}}^{k},\bA_{i}:Q_{i}^{k}
 }}
We write $C_{1}\Slar{\ga}_{\la_{\ga}}C_{k}$ where $\la_{\ga}=\prod_{j=1}^{k}\la_{j}$, and call $C_{k}$ a final configuration of $\cN$.
 \label{apath} 
 } 

Notice also that paths always terminate since event sequences are finite. 

One could straightforwardly define the operational semantics of a system $\cN$ to be the probabilistic transition system (PTS) defined by of all its paths.  However, we choose to identify those paths leading to the same observable behavior of an agent network.  Concretely, for particular inputs $\bi$ and $\rho \in I$, we identify final configurations for which only internal bindings in the local state of agents are different, that is, bindings for names not in the classical output set $\bo$.  These bindings correspond either to outcomes of measurements appearing in pattern events, or result from classical rendez-vous events.  As long as these are not part of the classical output, their actual values are unimportant. We cannot trace out these measurement outcomes after each pattern event, since a subsequent event may depend on these values.  Notice that the final sorts of agents \emph{do} need to be identical, as well as, obviously, the final network quantum state.  Identifying such paths, in the style of the measurement calculus, then gives us the semantics of a network of agents. However, because of the nondeterminism in the order in which concurrent agents execute their event sequence, we have to define the operational semantics of a network with respect to a particular \emph{schedule}. A schedule is precisely a particular order in which agents execute events. For example, in the network $\bA:\ens{1}.\cH(1,2)\nr\bB:\ens{3}.\cH(3,4)$, possible schedules are $\bA\bB$ and $\bB\bA$.  If we do not take schedules into account, we would add probabilities of all paths and all schedules resulting in identical final configurations, which for the example above leads to a probability of 2 for computing $H\ox H$ for arbitrary inputs, which is clearly not what we intend to say. We refrain from giving a formal definition for schedules, as we will find in \se{dsem} that the semantics of a network is independent of the schedule. Putting all this together, we obtain the following definition.

\DEF{ 
The \emph{operational semantics} of a network  $\cN=|_{i}\bA_{i}(\bi_{i},\bo_{i}):Q_{i}.\cE_{i}\given \sig $, with respect to a particular schedule, is a probabilistic transition system relating initial with final configurations, 
\EQ{
\ops \cN : \uplus_{i} Q_{i}\rar\uplus_{i} Q'_{i}. \ox_{i}\rho_{i},|_{i}\Ga_{\bi_{i}}\Lrar_{\la}\sig', |_{i}\Ga_{\bo_{i}}
 } 
with  $\la=\sum_{\ga}\la_{\ga}$ and the sum runs over all paths $\ga$ such that
\EQ{
\sig\ox_{i}\rho_{i},|_{i}(\Ga_{\bi_{i}},\bA_{i}(\bi_{i},\bo_{i}):Q_{i}.\cE_{i})\Slar{\ga}_{\la_{\ga}}\sig',|_{i}(\Ga_{\bo_{i}},\bA_{i}(\bi_{i},\bo_{i}):Q'_{i})\text{  .}
}
We call $\uplus_{i} Q_{i}\rar\uplus_{i} Q'_{i}$ the \emph{type} of the network.
\label{asemdef} 
}
From now on, we denote $I=\uplus_{i}I_{i}$ for the set of input qubits and $O=\uplus_{i} Q'_{i}$ for the set of output qubits, and call $\den{\cH_{I}}$ and $\den{\cH_{O}}$ the quantum input and output space respectively.
The semantics of a network with respect to a schedule is thus that it relates quantum states in $\den{\cH_{I}}$ plus classical input to quantum states in $\den{\cH_{O}}$ and classical output with particular probabilities. Note that the type of the transition system is a mapping from initial to final sorts; this component is identical in the denotational semantics we develop in \se{dsem}.

We say that two networks $\cN_{1}$ and $\cN_{2}$ are \emph{operationally equivalent} if their operational semantics, given by a PTS, is identical, and write $\ops{\cN_{1}}\equiv\ops{\cN_{2}}$.
In fact, we identify operational equivalence with the notion of \emph{bisimilarity}. This is indeed sensible since, by identifying computation paths as in \de{asemdef}, we actually impose a bisimilarity relation on final configurations. As we shall see in \se{televsqc}, it is by doing exactly this that we can show, among others, that teleportation is bisimilar to a direct quantum channel.

\subsection{Denotational semantics}\label{dsem}
Because of its more abstract character, in many situations it is more adequate to work with the denotational semantics. This is why we develop this notion for networks of agents. As before, it is closely related to the operational semantics, as well as to the semantics of ordinary patterns. Indeed, upon inspection of \de{asemdef}, we see that for any schedule, the PTS associated with a network of agents decomposes in several parts. First, there is a map from initial to final sorts, which determines how qubit ownership evolves for each of the agents from initial to final configurations. This is formalized as a type signature, exactly as we did for the operational semantics. The sort mapping is independent of the classical input: indeed, classical inputs appear only in classical communications, Pauli corrections and measurement angles, none of which affect qubit sorts. Furthermore, they can be read of statically from the network definition and are also schedule independent. The denotational semantics is then a mapping from classical inputs to classical outputs and a quantum operation, which in turn determines how quantum states evolve in the network. However, these two components are not independent of each other, since classical outputs can be measurement outcomes, which occur with probabilities that depend on the quantum operation applied. For simplicity, let us first consider the case where there are no classical outputs. In this case, for each classical input $\bi=\uplus_{i}\bi_{i}$ we have map $\cL$ which describes how the network quantum state evolves. This map is a multilocal quantum operation, because if we throw away all distributed information, that is, sorts and communication events, we just have an ordinary pattern, i.e. a quantum operation.  There is one caveat: since computation occurs asynchronously, there is usually some choice in the order in which different agents execute events in their program, i.e. there are different possible schedules. However, since at each instance of the computation local events operate on disjoint sets of qubits, it does not actually matter in which order these operations are applied, or, in fact, whether they are executed at the same time. This statement is proved formally in \se{contexts}; we postpone the full proof until then since it has bearing on other situations that are covered below. Therefore, any schedule of the computation leads to the same quantum operation. So to determine the operation elements of $\cL$, we choose a particular schedule, and then compose patterns in the order in which they are executed, tensoring with identity patterns where necessary and ignoring communication commands. Each operation element $L_{j}$ then corresponds to a sequential composition of actualizations for each of these patterns. 

Suppose now that the network contains classical outputs $\bo=\uplus_{i}\bo_{i}$. We need to make a distinction between \emph{signal} outputs, which are measurement outcomes, and \emph{external} outputs, which are values that were originally input by some agent and sent around the network. By definition, the external outputs $\bo_{e}=\uplus_{i}\bo_{i,e}$ depend only on the classical input \bi; these constant values are sent around the network via classical channels. It is precisely the signal outputs $\bo_{s}=\uplus_{i}\bo_{i,s}$ that depend on the quantum operation and vice versa. Indeed, when there are signal outputs particular measurement outcomes are preserved, therefore excluding actualizations of $\cL$ that do not correspond to that outcome. This essentially means that a different quantum operation is applied for each possible signal output. For example, suppose one of the signal outputs, corresponding to a measurement on qubit 3 is equal to $1$. Then only those actualizations containing the operator $\langle -_{\al}|_3$ are compatible with this output. We denote actualizations compatible with output $\bo_{s}$ by  $L_{i}^{\bo_{s}}$, and the quantum operation with these operation elements by $\cL^{\bo_{s}}$, and call these \emph{restricted}. Note, however, that this is a \emph{trace-decreasing} operation, and that $\tr \cL^{\bo_{s}(\rho)}$ is precisely the probability with which the output $\bo_{s}$ occurs. So, whereas the operational semantics gives explicit probabilities for each path, in the denotational semantics these are contained within the quantum operations. It is this abstraction, together with schedule-independence which makes the denotational framework advantageous. Indeed, classical inputs are the same for all schedules, and classical outputs, depending on classical inputs and measurement values, thus occur with the same probabilities for all schedules, since $\cL$ is schedule-independent. Putting all of this together, we arrive at the following definition.

\DEF{
The \emph{denotational semantics} of a network of agents $\cN=|_{i}\bA_{i}(\bi_{i},\bo_{i}):Q_{i}.\cE_{i}\given \sig$ is given by 
\EQ{
\des\cN:\uplus_{i}Q_{i}\rar \uplus_{i}Q'_{i}.\bi\rar\ens{(\bo,\cL^{\bo_{s}}), \forall \bo_{s}}
}
with
\EQ{
\cL: \den{\cH_{I}}\rar \den{\cH_{O}}:\ox_{i}\rho_{i}\rar  \sum_{j}L_{j}(\sig\ox_{i}\rho_{i})L_{j}^{\dag}\text{  ,}
}
where $\bo=\bo_{e}\uplus\bo_{s}$, $\rho_{i}$ is the quantum input, $Q'_{i}$ the final sort of agent $\bA_{i}$, and $I$ and $O$ are quantum input and output spaces respectively. In case there are no outputs, we have $\des\cN:\uplus_{i}Q_{i}\rar \uplus_{i}Q'_{i}.\bi \rar \cL$, or just $.\cL$ if there are no inputs either.
\label{adefdensem}}

Note that the $\bo_{e}$ part of $\bo$ in each of the above tuples is identical.
As an example, consider the pattern $\cx 1{s_{2}}\M{{-\al}}2$, which implements the bit-flip channel ~\cite{Nielsen00}. It can be interpreted as a one-agent network $\bA(-,-):\ens{1}.\cx 1{s_{2}}\M{{-\al}}2$, with has as denotational semantics the quantum operation $\cL(\rho)=p\rho+(1-p)X\rho X$. However, the one-agent network $\bA(-,\ens{s_{2}}):\ens{1}.\cx 1{s_{2}}\M{{-\al}}2$ has a different semantics, namely
\EQ{
\ens{(0,p\rho),(1,(1-p)X\rho X)}\text{  ,}
}
for all $\rho$, where $p$ is a function of $\al$. While this example may seem contrived, it is actually crucial that the semantics of these kind of networks are different, as they describe different states of  \emph{knowledge} of the output $s_{2}$, and hence, of what actual computation path  was taken. 

In case the underlying quantum operation $\cL$ is deterministic, all actualizations lead to the same quantum output. However, even in that case we require the above formulation with different trace-decreasing quantum operations $\cL^{\bo_{s}}$, since these determine the probabilities of outputting $\bo_{s}$. A network is deterministic only if for any input \bi, the quantum operation $\cL$  is deterministic, i.e. it implements a unitary, \emph{and} the classical output \bo is identical in all actualizations. 

As before, two networks are called denotationally equivalent if they have the same denotational semantics. Note that equivalent networks may have different preparation states, and also that agent names may be different, though the number of agents must be identical in both networks. With the above definitions in place, we can now prove the following result.

\PRO{There is a \emph{precise correspondence} between the operational and the denotational semantics of networks of agents, that is to say
\EQ{
\forall \cN_{1},\cN_{2}: \cN_{1}\equiv_{op}\cN_{2} \iff \cN_{1}\equiv_{de}\cN_{2} 
}
\label{aequivsem}}
\begin{proof}
Suppose $\des{\cN_{1}}=\des{\cN_{2}}$. This means that for all classical and quantum inputs, both the the type and classical external outputs are identical. Since for all signal outputs  $\cL^{\bo_{s}}_{1}= \cL^{\bo_{s}}_{2}$, we have that $\cL_{1}=\cL_{2}$ and therefore computation paths are also the same for both networks ~\cite{DHondt05}. Hence, both networks are operationally equivalent.
\end{proof}

Since both semantics' are equivalent, we can choose the operational or denotational framework at out convenience. We usually derive the denotational semantics via operational computation paths, relying also on the fact that the semantics is schedule independent.

\subsection{Compositionality}\label{acomp}

An important goal is to prove that the semantics of networks is conserved with respect to network composition as in \defs{netcomp}{netcompp}. Knowing this, we can compose any two networks and be ensured that the resulting network carries out the intended computation. In other words, we need to prove \pro{acompprop} below -- notice that a similar result exists for patterns~\cite{Danos04b}, though the operations are of course defined differently. However, in order to do this we require a proper definition for composing mathematical objects of the form of \de{adefdensem}. The reasonable way to this is to gather types, inputs and outputs, eliminating those that are fed from one network into the other in case of sequential composition, exactly like we did for agents in \de{agentcomp}. Next, we need to combine quantum operations by tensoring with the identity map were necessary. As we shall see below, in this way we already recover most of \defs{netcomp}{netcompp} . In fact, we only need to be careful in checking whether the quantum operations combine in the correct way, but, for this we can rely on the abovementioned \pro{acompprop}.

\PRO{
The semantics of networks is \emph{compositional}, i.e.
\AR{
\sem{\cN_{2}.\cN_{1}}&=\sem{\cN_{2}}.\sem{\cN_{1}}\\
\sem{\cN_{1}\ox\cN_{2}}&=\sem{\cN_{2}}\ox\sem{\cN_{1}}
}
\label{acompprop}
}
\begin{proof}
Suppose we have two networks $\cN_{1}=|_{i}\bA_{i}(\bi_{1,i},\bo_{1,i}):Q_{1,i}.\cE_{1,i}\given \sig_{1}$ and $\cN_{2}=|_{i}\bA_{i}(\bi_{2,i},\bo_{2,i}):Q_{2,i}.\cE_{2,i}\given \sig_{2}$ with semantics given by
\EQ{\SP{
\sem{\cN_{1}}&:\uplus_{i}Q_{1,i}\rar \uplus_{i}Q'_{1,i}.\bi_{1}\rar\ens{(\bo_{1},\cL^{\bo_{s}}_{1})}\\
\sem{\cN_{2}}&:\uplus_{i}Q_{2,i}\rar \uplus_{i}Q'_{2,i}.\bi_{2}\rar\ens{(\bo_{2},\cL^{\bo_{s}}_{2})}\text{  ,}
}}
with 
\EQ{\SP{
\cL_{1}&: \den{\cH_{I_{1}}}\rar \den{\cH_{O_{1}}}\ox_{i}\rho_{1,i}\rar \sum_{j}L_{1,j}(\sig_{1}\ox_{i}\rho_{1,i})L_{1,j}^{\dag}\\
\cL_{2}&: \den{\cH_{I_{2}}}\rar \den{\cH_{O_{2}}}\ox_{i}\rho_{2,i}\rar \sum_{j}L_{2,j}(\sig_{2}\ox_{i}\rho_{2,i})L_{2,j}^{\dag}\text{  .}
}}
We then find that
\EQ{
\sem{\cN_{2}}.\sem{\cN_{1}}:\uplus_{i}Q_{i}\rar \uplus_{i}Q'_{i}.\bi \rar \ens{(\bo,\cL^{\bo_{2,s}}_{2}.(\cL^{\bo_{1,s}}_{1}\ox \cI)), \forall \bo_{1,s},\bo_{2,s}}\\
}
where type, classical input and output are found by pointwise application of the rules in \eq{typecomp}, and $\cI$ is the identity operation on $I_{2}\bs O_{2}$. Quantum operations in the above map states of the form $\ox_{i}\rho_{1,i}\ox_{i}\rho_{2,i}$ in  $I=I_{1}\uplus(I_{2}\bs O_{2})$ to states in $O=O_{2}$, 
after tensoring them with $\sig_{1}\ox\sig_{2}$;  On the other hand, the semantics of $\cN_{2}\circ \cN_{1}$ is given by 
\EQ{
\sem{\cN_{2}\circ \cN_{1}}: \uplus_{i}Q_{i}\rar \uplus_{i}Q'_{i}.\bi \rar \ens{(\bo,[\cL_{2}.(\cL_{1}\ox \cI)]^{\bo_{s}}), \forall \bo_{s}}\text{  ,}
}
where $\bo_{s}=\bo_{1,s}\uplus\bo_{2,s}$, and the quantum operations operate on the same $I$ and $O$ as above. 
So we only need to check that first composing the quantum operations and then restricting them is the same as first restricting and then composing. However, this follows from the analogous result for ordinary patterns~\cite{Danos04b}.

For parallel composition, consider instead $\cN_{2}=|_{i=1}^{n}\bB_{i}(\bi_{2,i},\bo_{2,i}):Q_{2,i}.\cE_{2,i}\given \sig_{2}$, with semantics as above. We then find that
\EQ{
\sem{\cN_{1}}\ox\sem{\cN_{2}}:\uplus_{i}Q_{i}\rar \uplus_{i}Q'_{i}.\bi \rar \ens{(\bo,\cL^{\bo_{1,s}}_{1}\ox\cL^{\bo_{2,s}}_{2}), \forall \bo_{1,s},\bo_{2,s}}\text{  ,}
}
where types, inputs and outputs are found by taking the (disjoint) union of those of the composing networks. Again, via the analogous result for ordinary patterns~\cite{Danos04b}, we find that $(\cL_{1}\ox\cL_{2})^{\bo_{s}}=\cL^{\bo_{1,s}}_{1}\ox\cL^{\bo_{2,s}}_{2}$, and therefore it follows that the above expression equals $\sem{\cN_{1}\ox\cN_{2}}$.
\end{proof}

\subsection{Entanglement contexts}\label{contexts}

In the previous section,  the networks $\cN_{1}$ and $\cN_{2}$ individually operate on tensor product states of the form  $\ox_{i}\rho_{i}$, in accordance with the fact that these are local inputs provided by each of the agents, as is made explicit in \de{asemdef}. While this is sensible when considering a network operating in isolation -- in which case entangled input states are specified as preparations -- it is less so when a network is only one factor in a complex compositional structure. Indeed, already for sequential composition the input state is in general no longer disentangled over agents, since it is fed in partly as output of a previous network computation, and there is no guarantee whatsoever that this output is a product state. Another subtle difference lies in the fact that input spaces of individual agents are combined when composing networks. Concretely, while each agent supplies local inputs $\rho_{1}$ when $\cN_{1}$ is run separately, and $\rho_{2}$ when $\cN_{2}$ is run separately, inputs to the composed network $\cN_{2}\circ\cN_{1}$ are generally \emph{not} of the form $\rho_{1}\ox\rho_{2}$. Yet another situation is that where agents have mixed state inputs because their inputs are entangled into a state on a larger system than that which the network operates on --  in fact one view is that this is the \emph{only} way in which mixed states arise. These are of course all typical manifestations of entangled states, but we want to stress the different situations in the context of distributed networks in which these arise. The point is, is entanglement preserved when applying operations to only part of the entangled state? We show below that it does, by proving the a quantum operation applied to the $A$-system of a state $\rho$  existing on system $AC$ does not touch the part of this state in $C$. Each of the aforementioned situations can be cast in this form, since $A$ may be a system of several agents entangled with a group of other agents described by system $C$, as well as an input system of one agent only entangled with another input system $C$. The quantum operation itself is defined to map states on $A$ to states on $B$, because we want to consider situations where a network's output spaces is different than its input space. 

\PRO{
Suppose $\cL$ is a quantum operation on system $A$ to system $B$ such that
\EQ{
\cL:\den{\cH_{A}}\lrar \den{\cH_{B}}:\rho_{A}\lrar\rho_{B}=\sum_{k}L_{k}\rho_{A}L_{k}^{\dag}\text{  .}
}
Then for all quantum states $\rho_{AC}$ living on a system $AC$, applying $\cL$ to one half of $\rho_{AC}$ results in
\EQ{
\rho_{BC}=\sum_{k}(L_{k}\ox I_{C})\rho_{AC}(L_{k}^{\dag}\ox I_{C})
}
\label{entcont}}
\begin{proof}
Note first that any complex matrix can be written as a linear combination of Hermitian matrices, which in turn can be written as a sum of density matrices.  Then by the spectral decomposition and linearity, it follows that for all complex matrices $Z$ we have $\cL(Z)=\sum_{k}L_{k}Z L_{k}^{\dag}$.  Writing $\rho_{AC}=\sum_{ijkl}\al_{ijkl}\ket{i}_{A}\ket{j}_{C}\bra{k}_{A}\bra{l}_{C}$, we find that

 \EQ{\SP{ 
(\cN\ox I_{C})\rho_{AC}&=\sum_{ijkl}\al_{ijkl}(\cN\ox I_{C})(\ketbra{i}{k}^{A}\ox\ketbra{j}{l}^{C})\\ 
&=\sum_{ijkl}\al_{ijkl}(\sum_{k}(L_{k}\ketbra{i}{k}^{A}L_{k}^{\dag})\ox\ketbra{j}{l}^{C})\\ 
&=\sum_{k}(L_{k}\ox I_{C})\rho_{AC}(L_{k}^{\dag}\ox I_{C}) \text{  ,} }
 } 
which proves the theorem.  
\end{proof}
 
The proof, while easy, is not trivial, and has several important consequences. First of all, it shows that our statement in \se{dsem} on the fact that networks are schedule-independent is true. Indeed, agents transform parts of a shared entangled state via local operations, and the above results shows that any order results in the same multilocal quantum operation. Next, while we have proved compositionality in the previous section considering only product state, we are now ensured that it holds also for arbitrary input states. We rely on \pro{entcont}  also in the next section, when we discuss sequencing several teleportation networks to transfer an entangled state from one agent to another.

\section{Teleportation vs. quantum channels}\label{televsqc}

In this section we prove that the teleportation protocol, described within our framework of networks of agents, is bisimilar to a direct quantum communication of the qubit to be teleported. We first give the network specification for direct quantum communication. Next, we give the network for the teleportation protocol, rederiving the correctness of the protocol by developing its semantics in the local view elaborated in \se{defs}, which gives us a handle on qubit locations. A second goal of this section is to prove that it is bisimilar to a direct quantum channel. That is, by comparing  the semantics of a direct quantum channel and that of teleportation -- evolved in the correctness proof --  we conclude that they do indeed define identical PTS's.  By \pro{entcont}, this holds in arbitrary entanglement contexts. This is a nontrivial result because these agents may be entangled with the qubit to be teleported, and it is not a priori clear whether this entanglement is preserved throughout TP.  This result has several consequences that we mention below.  While the correctness of teleportation has been proved within other formal frameworks for distributed systems before~\cite{Gay04,Abramsky04}, the bisimilarity approach, and specifically, taking into account arbitrary contexts, is new. 

Consider a network of two agents named \bA and \bB. A direct quantum communication of a qubit is implemented simply by the network

\EQ{
\cN=\bA:\ens{1}.(\qc!1) \nr \bB.(\qc?1)\given  \mbf{0} \text{  ,} 
}
where $\mbf{0}$ is the null state. Note that \bB has an empty sort. The small-step semantics, given input $\gket$, is derived in one step, and leads to the operational semantics

 \EQ{ 
 \sem {\cN}: (\ens{1},-)\rar(-,\ens{1}).\gket \Lrar \gket \text{  ,}
}
which immediately is the denotational semantics as well.

Consider the following network definition, which, as we derive explicitly below, implements the teleportation protocol. 

\EQ{\SP{ 
\bfa&=\bA:\ens{1,2}.[(\cc!s_{2}s_{1}).\tM 0012]\\ 
\bfb&=\bB:\ens{3}.[\cx 3{x_{2}}\cz 3{x_{1}}.(\cc?x_{2}x_{1})]\\ 
\cN_{TP}&=\bfa \nr \bfb \given \et 23}
}

To derive the semantics of the above network, note that for the first step there is only one possibility, namely that agent $\bA$ executes the local Bell measurement $\tM 0012$. The latter requires a local quantum input from $\bA$, namely the qubit $\gket$ that needs to be teleported.  This is clearly the case since the pattern applies to qubit $1$, which is not part of the system state $ \et 23$.  So by rule \eqref{localop} we need to apply the pattern to the first two qubits of $\gket\ox  \et 23$.  Using first rule \eqref{localop}and writing $\Ga_{\bfa}=\emptyset[s_{2}s_{1}\mapsto j_{2}j_{1}]$, we derive

\EQ{ 
\frac{\gket_{1}\ox  \et 23,\tM 0012\lrar_{1/4} X^{j_{2}}Z^{j_{1}}\gket_{3}, \Ga_{\bfa}}
{ \et 23,\emptyset,\bfa\Lrar_{1/4} X^{j_{2}}Z^{j_{1}}\gket_{3}, \Ga_{\bfa},\bA.(\cc!s_{2}s_{1})} \text{  ,}
}

where for each of the values of $j_{2}j_{1}$ the transition occurs with the same probability of $1/4$. This reduction fires within the context of  \eqref{metarule}, which we do not write out explicitly. The next step is a classical rendez-vous between both agents (i.e.  Alice calls Bob), as per rule \eqref{rendvcl}.  Defining $\Ga_{\bfb}=\emptyset[x_{2}x_{1}\mapsto j_{2}j_{1}]$, we get

\EQ{\SP{ 
X^{j_{2}}Z^{j_{1}}\gket_3 \ent (\Ga_{\bfa},\bA&.(\cc!s_{2}s_{1})\nr\emptyset,\bB:\ens{3}.[\cx 3{x_{2}}\cz 3{x_{1}}.(\cc?x_{2}x_{1})]\\ 
&\Lrar  \Ga_{\bfa},\bA\nr \Ga_{\bfb},\bB:\ens{3}.\cx 3{x_{2}}\cz 3{x_{1}})\text{  .} }
}

The last step of the computation is the execution of a local pattern by agent $\bB$, as follows.

\EQ{ 
X^{j_{2}}Z^{j_{1}}\gket_3, \Ga_{\bfa},\bA\nr \Ga_{\bfb},\bB:\ens{3}.\cx 3{x_{2}}\cz 3{x_{1}} \Lrar \gket_{3},\Ga_{\bfa},\bA\nr \Ga_{\bfb},\bB:\ens{3}
}

The only probabilistic transition is the first one, due to the Bell measurement.  However, we see that the four branches lead to identical final system state and agents specifications.  Furthermore, since there is no classical input or output, we can trace out the different local states.  Thus, adding the probabilities as specified in \de{asemdef},  we find that for any input $\gket$, we have 

\EQ{
 \sem \cN_{TP}: (\ens{1,2},\ens{3})\rar(-,\ens{3}).\gket \Lrar \gket 
 } 
 
In other words, we find that $\sem{\cN_{TP}}\equiv \sem{\cN}$.

Note that by linearity, the above derivation also works for mixed states.  

We have shown that a direct quantum channel is operationally equivalent to the teleportation protocol.  Suppose however, that two agents wish to exchange a qubit whilst they are contained in a larger network of agents.  Can we say anything about the equivalence of both procedures in this context? By \pro{entcont}, we know that the entanglement with the larger system is conserved. Indeed, since teleportation just implements an identity channel, applying teleportation to one half of the mixed state $\rho_{AC}$ results in the state $\rho_{BC}$. Furthermore, suppose agent \bA wants to send an $n$-qubit entangled state to \bB. Then by the same result, \bA can just apply the teleportation protocol $n$ times, and, since entanglement is conserved, the state is transferred unchanged. It remains to be seen whether the conservation of correlations between agents can also be employed in higher-level applications. Considering the fact that shared entanglement provides much of the extra power in distributed quantum systems, this behavior seems a promising primitive.

\section{Conclusion} \label{aconclusion}
In this paper, we develop a formal model for distributed measurement-based quantum computations. We adopt an agent-based view, such that computations are described locally where possible. Because the network quantum state is in general entangled, we need to model it as a global structure, reminiscent of global memory in classical agent systems. Local quantum computations are described as measurement patterns. Since measurement-based quantum computation is inherently distributed, this allows us to extend naturally several concepts of the measurement calculus (MC)~\cite{Danos04b}, a formal model for such computations. Just as in MC, we aim at defining an \emph{assembly language}, i.e. we assume that computations are well-defined  and do not concern ourselves with verification techniques. The  operational semantics for systems of agents is given by a probabilistic transition system, and we define operational equivalence in a way that it corresponds to the notion of \emph{bisimilarity}. The denotational semantics is given by a set of quantum operations, together with type information which determines the localization of resources. Both forms of semantics are proved to be equivalent, and we define a notion of network composition such that the semantics is preserved with respect to this operation. Moreover, we show that within the larger entanglement contexts, the semantics is also conserved. With this in place, we prove that teleportation is bisimilar to a direct quantum channel, and, by the abovementioned result, this also holds within the context of a larger network. Though the proof is quite simple, it is important within the context of agent systems. Indeed, the possibility of inheriting agent correlations via teleportation means that, for example, collaborating agents are not cut off from each other when part of the shared data is transferred.  Rather, the correlations are preserved in an oblivious manner.  That this is the case if qubits are transported physically is clear, but that this remains so even if a general protocol is employed is maybe more surprising. It remains to be investigated how this preservation of the entanglement context can be exploited in more general situations.  Also, there are other interesting situations to be investigated, such as the so-called \emph{channel inequalities}, and more elaborate communication protocols. This is the subject of current investigations.

\bibliographystyle{alpha} 
\newcommand{\etalchar}[1]{$^{#1}$}

\end{document}